\documentstyle[preprint,aps,prl]{revtex}
\begin{document}
\bibliographystyle{prsty}
\draft
\title{The Spin Stiffness and the Transverse Susceptibility of the
Half-filled Hubbard Model}
\author{Zhu-Pei Shi and Rajiv R. P. Singh}
\address{Department of Physics, University of California, Davis, California
9561
6}
\maketitle
\begin{abstract}
The $T=0$ spin stiffness $\rho _{s}$ and the transverse susceptibility
$\chi _{\perp }$ of the square lattice half-filled
Hubbard model are calculated as a function of the Hubbard parameter
ratio $U/t$ by series
expansions around the Ising limit. We find that the calculated spin-stiffness,
transverse susceptibility, and sublattice magnetization
for the Hubbard model smoothly approach
the Heisenberg values for large $U/t$. The results are compared for different
$U/t$ with RPA and other numerical studies.
The uniform susceptibility data indicate a crossover around $U/t\approx 4$
between weak coupling (spin density wave)
behavior at small $U$ and strong coupling ( Heisenberg ) behavior at large $U$.

\end{abstract}

\narrowtext

Recent discovery of
high-$T_{c}$ superconductivity in the Cuprate materials
has generated tremendous interest
in the subject of strongly correlated electrons. In these systems,
the phenomena of
unusual metallic behavior, antiferromagnetism and
superconductivity
occur in a narrow parameter range.
It is widely believed that these phenomena have a common microscopic
origin.
The Hubbard model is one of the simplest models to describe correlated electron
behavior in a solid.
It consists of a single
band of electrons, with nearest neighbor hopping parameter
$t$ and an on-site Coulomb repulsion between opposite spin electrons of
magnitude  $U$. The model is best understood at half-filling, that is,
when there is one electron per unit cell,
where the system becomes an antiferromagnetic insulator.
At large values of $U$ this system is well described by the Heisenberg
model. At small $U$
one expects the spin-density-wave (SDW) mean field description to
become accurate.
The possibility of $d$-wave superconductivity, away from half-filling,
has been widely explored~\cite{dog:95}.

Direct calculations of the superconducting transition temperatures
in the Hubbard model are beyond present numerical capabilities.
Thus, phenomenological approaches, where the Hubbard model is used
to determine the parameters in a scaling theory of superconductivity,
are appropriate.
If spin-fluctuations are important
to the mechanism of superconductivity in the Cuprates,
and if the magnetic excitations in the doped Cuprates
are related to those in the stoichiometric insulating phases,
the magnetic excitations in the half-filled system are clearly important for
understanding superconductivity in these materials.

The magnetic ground state of the
two-dimensional (2D) square lattice Hubbard model
has been investigated by quantum Monte Carlo
simulations~\cite{hir:89,whi:89,mor:90} and Lanczos
diagonalization~\cite{dag:89}. These studies have mainly consisted of
a finite size scaling analysis of the ground state properties of the model.
They confirm the existence of long-range
antiferromagnetic order at $T=0$. To our knowledge, they have not been
used to calculate the spin-stiffness constant or the spin-wave velocity
for this model. Accurate numerical calculations of these quantities
exist for the Heisenberg model.
The Spin-density wave theory combined with the
random phase approximation (RPA) has been used by
Schrieffer et al.~\cite{rpa:89} to
calculate the spin wave velocity for the Hubbard model. This calculation
should become exact as $U\to 0$.
Somewhat surprizingly, it was found that
the result of this  calculation is also accurate
in the Heisenberg limit
($U/t \gg 1$).

In this Letter we {\em first\/} derive an expression for the
spin stiffness constant of the Hubbard model by applying a slow twist in
the ordering direction. We {\em then} introduce a one-parameter
family of Hamiltonians which interpolate between the half-filled
Hubbard and Ising models. This allows us to develop series expansions
for the spin-stiffness.
In addition, we develop series expansions for several
thermodynamic parameters of this model, such as the
uniform susceptibility and the sublattice magnetization.
The spin wave velocity is calculated
from the hydrodynamic relation~\cite{hal:69} $v_s^2 =\rho_s/\chi_\perp$.

At large $U$ our results extrapolate smoothly to the Heisenberg values.
They also show good agreement with the spin-density wave theory at small $U$.
The variation of the magnetic susceptibility with the Hubbard
parameter ratio $U/t$ shows a relatively well defined crossover
between a $\chi _{\perp }\sim U$ behavior at large $U$ and $\chi_{\perp }$
decreasing with increase in $U$ at small $U$. This crossover between
the strong coupling ( Heisenberg model behavior) and weak coupling SDW
behavior occurs at
$U/t\approx 4$.

The Hubbard model is defined by the lattice Hamiltonian
\begin{equation}
H_{0}=-t\;\sum _{<i,j>,\sigma }\; (c^{+}_{i\sigma }\,c_{j\sigma }
                               +c^{+}_{j\sigma }\,c_{i\sigma })
  +U\sum _{i}\;(n_{i\uparrow }-\frac{1}{2})\,(n_{i\downarrow }-\frac{1}{2})
\;-\; \mu \sum _{i} \;(n_{i\uparrow }+n_{i\downarrow })\;\;\; ,
\end{equation}
where $c^{+}_{i\sigma }$ and $c_{i\sigma }$ are the creation and
annihilation operators for electrons with a $z-$component of
spin $\sigma $ at lattice site $i$,
and $n_{i,\sigma }=c^{+}_{i\sigma }\,c_{i\sigma }$.
$U$ is the on site repulsive interaction, $\mu $ the chemical potential,
and $t$ the nearest-neighbor hopping amplitude.

If we rotate the ordering direction by an angle $\theta $ along a
given direction such as $y$ axis, then the spin stiffness constant
$\rho _{s}$ can be defined through the increase of the ground state energy:
$
E_{g}(\theta )=E_{g}(\theta =0)\;+\; \frac{1}{2}\, \rho _{s}\,\theta ^{2}\;
+\; O(\theta ^{4}).
$
This rotation can be carried out by the following transformation applied to
the fermion operators:
\begin{equation}
\left(
\begin{array}{c}
c'_{\uparrow }\\
c'_{\downarrow}
\end{array}\right)\; =\;
\left(\begin{array}{cc}
cos\,\phi & sin\,\phi \\
-sin\,\phi  & cos\,\phi
\end{array}\right)\;\;
\left(
\begin{array}{c}
c_{\uparrow }\\
c_{\downarrow}
\end{array}\right)
\;\;\; .
\end{equation}

After rotation by a {\em relative\/} angle $\theta $
( that is letting $\phi$ change by $\theta/2$ ) between neighboring
sites separated along $y$ axis ($\hat y$ denotes unit distance in $y$
direction),
$H_{0}$ in Eq.(1) becomes $
H=H_{0}\;+\; H^{dia}\;+\; H^{para} \;+\;O(\theta ^{3}),
$
where
\begin{eqnarray}
H^{dia}&=&\frac{t\theta ^{2}}{8}\;\sum _{i,\sigma }\; (c^{+}_{i\sigma }
                                  \,c_{i+\hat{y}\sigma }
                               +c^{+}_{i+\hat{y}\sigma }\,c_{i\sigma
})\nonumber\\
H^{para}&=& -\frac{t\theta }{2}\;\sum _{i}\; (c^{+}_{i\uparrow }
\,c_{i+\hat{y}\downarrow }\;-\; c^{+}_{i\downarrow }c_{i+\hat{y}\uparrow }
   \;+\; c^{+}_{i+\hat{y}\downarrow } c_{i\uparrow }\;-\;
c^{+}_{i+\hat{y}\uparrow }c_{i\downarrow })\;\;\; .
\nonumber
\end{eqnarray}
The ``diamagnetic" term $H^{dia}$ is already of order $\theta ^{2}$ so
for the calculation of the energy to order $\theta^2$, it
can be replaced by its expectation value in the ground state of the $\theta =0$
Hamiltonian. We get
$
\rho ^{dia}_{s}=-\frac{1}{8}\; (Kinetic~ Energy)
  =-\frac{1}{8}\;[E_{g}(\theta =0)-\frac{U}{2}\;(n-L)] ,
$
where $n$ is a band filling, and $L$ the local moment defined as
$L= <(n_{i\uparrow }-n_{i\downarrow })^{2}>$.
The contribution of the ``paramagnetic" term $H^{para}$ to the ground state
energy in order $\theta ^{2}$ can be obtained
from the expression,
$\rho ^{para}_{s}=2\frac{\partial ^{2}E}{\partial \theta ^{2}}\mid _{\theta
=0},
$
where $E$ is the  energy of the Hamiltonian $H_{0}+H^{para}$.

In order to calculated these quantities numerically, we
introduce an Ising
anisotropy into the Hubbard Hamiltonian:
\begin{equation}
H_{0\lambda }=-\lambda \,t\;\sum _{<i,j>,\sigma }\; (c^{+}_{i\sigma }
                               \,c_{j\sigma }
                               +c^{+}_{j\sigma }\,c_{i\sigma })
  +U\sum _{i}\;(n_{i\uparrow }-\frac{1}{2})\,(n_{i\downarrow }-\frac{1}{2})
\;+J(1-\lambda ) \;\sum _{<i,j>}\;\sigma ^{z}_{i}\, \sigma ^{z}_{j}\;\;\;\; ,
\end{equation}
where $\sigma ^{z}_{i}=(n_{i\uparrow }-n_{i\downarrow })$ is the $z$ component
of the spin at site $i$,
and $J$ a parameter which can be tuned to improve the convergence of
the extrapolations. The particle-hole symmetry ensures half-filling.
For $\lambda =0$ the atomic limit of the Hubbard model is highly degenerate,
however the Ising term selects from these the N\'eel states as the two
degenerate ground states. Futhermore, this term also introduces a gap in
the spectrum at $\lambda =0$.
{\em For $\lambda =1$ the Ising anisotropy goes to zero and the conventional
Hubbard Hamiltonian is recovered\/}. Ground state properties of the model
for $\lambda \neq 0$ can be obtained by an expansion in powers of $\lambda $.
If the gap does not close before $\lambda =1$ as expected for this model, we
can obtain properties of the Hubbard model by extrapolating the expansions to
$\lambda =1$.
In the strong coupling limit, the half-filled Hubbard model is equivalent
to the Heisenberg model with Hamiltonian
$-J_{H}\; \sum _{<i,j>}\; (\sigma _{i}\cdot \sigma _{j}\;-\; 1)\;\;
$ with
$J_{H}=t^{2}/U$.
The optimum value of the parameter $J$ is found to be near
$J_{H}$~\cite{shi:95}.

We calculate series coefficients for the ground state energy
and the local moment to
$11th $ order, and $\rho ^{para}$ to $9th$ order in $\lambda $.
Using the Pad\'e analysis, we obtain spin stiffness
$\rho _{s}=0.186(15), 0.15(1), 0.077(7)$ and $0.039(5)$ for
$U=1, 4, 10$ and $20$, respectively.
They are plotted in Fig.~1 as filled circles with errorbars. The dashed line
is a  guide to the eye.
The errorbars represent the spread in the Pad\'e estimates.
At $U=20$, our result ($\rho _{s}U=0.78$) can be compared to that ($0.73$)
of the Heisenberg
model~\cite{sin:89}. Our results for $\rho_s$
as well as $M^\dagger$ are somewhat higher than the known results for
the Heisenberg models. We believe this reflects the fact that the
series have not converged as well as for the Heisenberg model and hence
the reduction in these quantities due to the zero-point
spin-wave fluctuations
is not fully accounted for. Still, the convergence ($\sim 5\% $)
is quite reasonable.

We also compare our results with the
Hartree-Fock approximation for the spin stiffness:
$
\rho _{s}=-\frac{t}{2N}\;\sum _{{\bf k}}\; \frac{\epsilon ({\bf k})\;
           cos\,k_{x}}{E({\bf k})} ,
$
where $\epsilon ({\bf k})=-2t\,(cos\,k_{x}+cos\,k_{y})$ and
$E({\bf k})=[\epsilon ^{2}({\bf k})+\Delta ^{2}] ^{1/2}$. $\Delta $ can
be obtained by solving the gap equation
$
\frac{U}{2N}\; \sum _{{\bf k}} \; \frac{1}{(\epsilon ^{2}_{{\bf k}}
                    +\Delta ^{2})^{1/2}}\;=\; 1.
$
The spin stiffness given by mean field solution of the gap equation is
plotted in Fig.~1 as a solid line. One can see that this approximation
overestimates the stiffness at large $U$.
At $U=20$, the mean field result is $\rho _{s}U=0.98$ compared to
$0.73$ for the Heisenberg model.

We note that the spin stiffness (filled squares in Fig.~1) from
the variational Monte Carlo method
with a Gutzwiller-type wave function are even larger than values of
the mean field solution~\cite{den:93}.
We believe that the large discrepancy is due to the missing spin flip
processes in the Gutzwiller variational wave function used in ~\cite{den:93}.
Their calculations only get contributions to the spin-stiffness
beyond the Hartree-Fock result from the ``diamagnetic'' term,
whereas the ``paramagnetic" part of the spin stiffness which contains
spin-flip processes does not get corrected~\cite{note:95}.

The transverse susceptibility of the Hubbard model can be
defined
in the usual way as
$
4\chi _{\perp }=-2\frac{\partial ^{2}E}{\partial h^{2}}\mid _{h=0} ,
$
where $E$ is the ground state energy of the Hamiltoniam
$H_{0\lambda }-h\,\sum _{i}\;\sigma ^{x}_{i}$.
The expansion coeffients of $4\chi _{\perp }$ to $9th$ order
in $\lambda $ are also calculated.

The transverse susceptibility obtained from Pad\'e approximants are
$4\chi _{\perp }=0.94, 0.58, 0.75, 0.92$ and $1.40$ for
$U=1, 4, 6, 10$ and $20$, respectively.
They are plotted in Fig.~2 as filled circles. A dashed line  is
a guide to the eye.
At $U=20$, $4\chi _{\perp }/U=0.07$ is close to the
large $U$
 limit value $\chi _{\perp }\,J'=0.065$ for
the Heisenberg  model~\cite{sin:89}
where $J'=4t^{2}/U$.
At $U=4$, our result ($4\chi _{\perp }=0.58$) also agrees with the
Monte Carlo Simulations ($\approx 0.53$)
by White {\em et al.\/}~\cite{whi:89}.
At $U=10$, our result ($4\chi _{\perp }=0.92$) agrees with the
Monte Carlo simulations of Moreo~\cite{mor:90} ($0.98$).
One can see in Fig.~2. that the variation of the magnetic susceptibility
with $U$ changes character around
$U=U_{c}\approx 4$.
It suggestes that there is a crossover at $U_{c}$ in the
behavior of the 2D Hubbard model.
For $U>U_{c}$, the magnetic susceptibility ($4\chi _{\perp }$) is
roughly proportional to $U$ or inversely proportional to the spin
superexchange $J$.
This can be interpreted to imply
that the magnetic state for $U>U_{c}$ for 2D Hubbard model is close
 to the Heisenberg model. But, for
$U<U_{c}$, the magnetic susceptibility ($4\chi _{\perp }$)
has qualitatively different behavior. It
decreases as $U$ increases. This can be interpreted as the weak
coupling behavior of the SDW antiferromagnetic state.

We also compare our result with the mean field solutions of uniform magnetic
susceptibility:
$
\chi _{\perp }=\frac{1}{2U}\;[(1-\Delta
^{2}\,U\,I_{1}-U\,I_{2}^{2}/I_{1})^{-1}\;-\;1] ,
$
where
$
I_{1}=\frac{1}{2N}\; \sum _{{\bf k}}\; \frac{1}{E^{3}({\bf k})}$
and $
I_{2}=-\frac{1}{2N}\; \sum _{{\bf k}}\; \frac{\epsilon ({\bf k})}{E^{3}({\bf
k})} $.
It is plotted in Fig.~2 as a solid line. Again, we see that
the mean field result is
qualitatively correct, but overestimates the quantity especially at large $U$.

The spin wave velocity obtained from the relationship
($\rho _{s}=v^{2}_{s}\chi _{\perp }$) are
$v_{s}=0.89,~1.02,~0.58$ and $0.34$ for
$U=1,~4,~10$ and $20$, respectively.
They are plotted in Fig.~3 as filled circles with a solid line
as a guide to the eye. At $U=20$, our $v_{s}=0.34$ agrees well with
the limiting value ($v_{s}=1.18\sqrt{2}\,J'=0.33$) of
the Heisenberg model~\cite{sin:89}.

We compare this result with the RPA solution of
Schrieffer, Wen and Zhang~\cite{rpa:89}:
$
v_{s}=[4t^{2}(1-\Delta ^{2}U\,I_{1})\;\frac{I_{3}}{I_{1}}]^{1/2},
$
where
$
I_{3}=\frac{1}{2N}\; \sum _{{\bf k}}\; \frac{sin^{2}\,(k_{x})}{E^{3}({\bf k})}
{}.
$
It is plotted in Fig.~3 as a dotted line. We see that the RPA result for the
spin
wave velocity are lower than our series expansion results.

To summarize, in this paper
we have derived an expression for the spin stiffness of the
Hubbard model, and calculated it and other key long-wavelength
parameters by systematic series expansion methods.
Comparison with the 2D $S-1/2$ Heisenberg antiferromagnet and RPA results,
suggests that our extrapolations are good at small and large $U$.
The uniform magnetic susceptibility data suggests a crossover between
SDW behavior at small $U$ and Heisenberg behavior at large $U$
around $U/t\approx 4$.

This research was supported in part by NSF (DMR-9318537) and
University Research Funds of the University of California at Davis.
We would like to thank Prof. R. T. Scalettar and
Prof. B. M. Klein for discussions.

\pagebreak
\begin{center}
        Figure Captions
\end{center}

Fig.~1 \hspace{0.6cm}
The spin stiffness as a function of $U$. Filled circles are our calculated
data, with a dashed line as a guide to the eye. The errorbars indicate
spread of the Pad\'e approximants.
The solid line is the mean field result.
The filled squares are results of variational Monte Carlo
simulations~\cite{den:93}.

Fig.~2 \hspace*{0.6cm}
Transverse  susceptibility as a function of $U$.
Filled circles are our calculated data, with a dashed line as a guide to the
eye.
A crossover between SDW and Heisenberg
antiferromagnet is suggested  around $U\approx 4$.

Fig.~3 \hspace*{0.6cm}
The spin wave velocity as a function of $U$. Filled circles are our calculated
data, with a solid line as a guide to the eye. The dotted line and dashed line
represent results of RPA and Heisenberg antiferromagnet respectively.
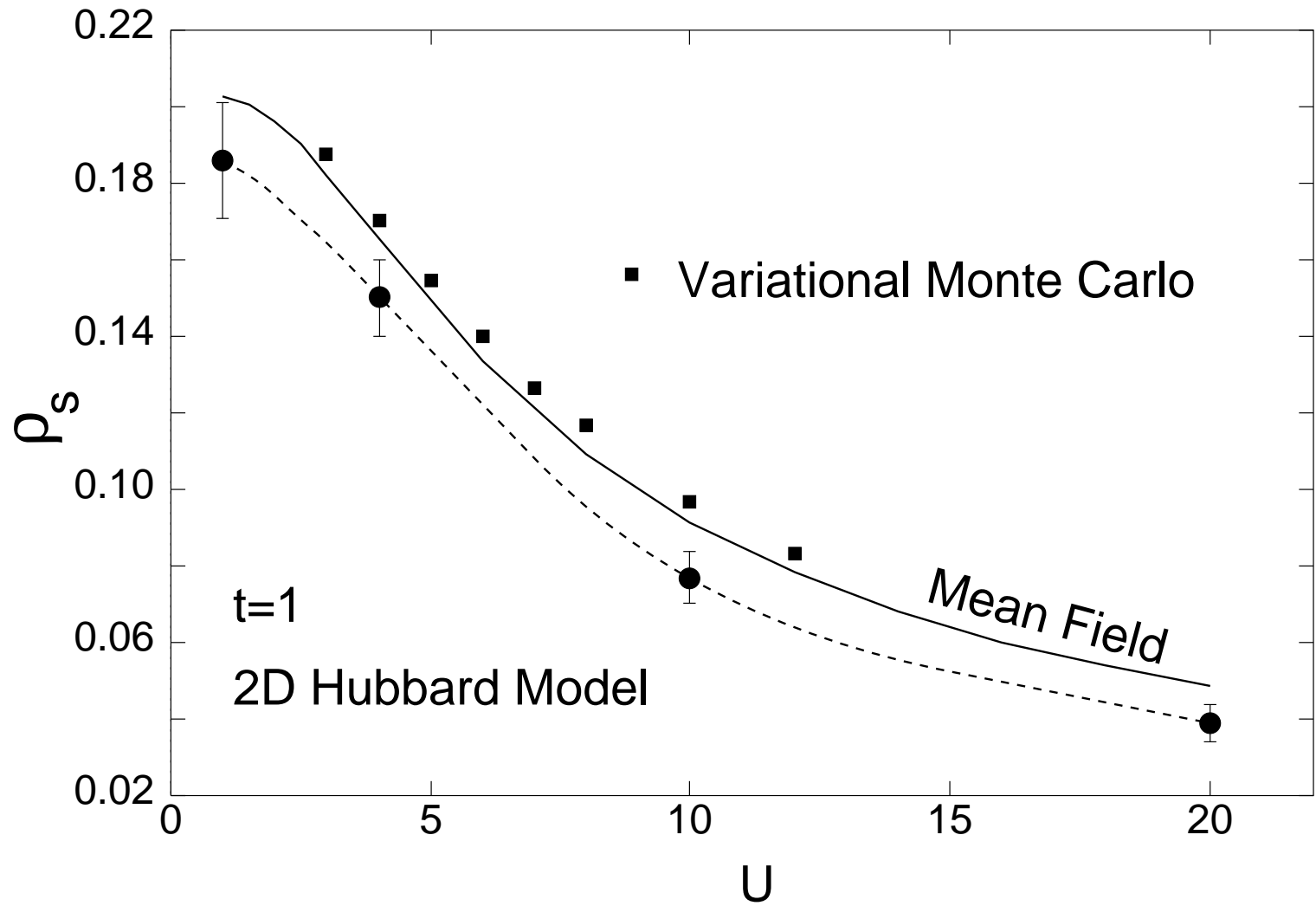
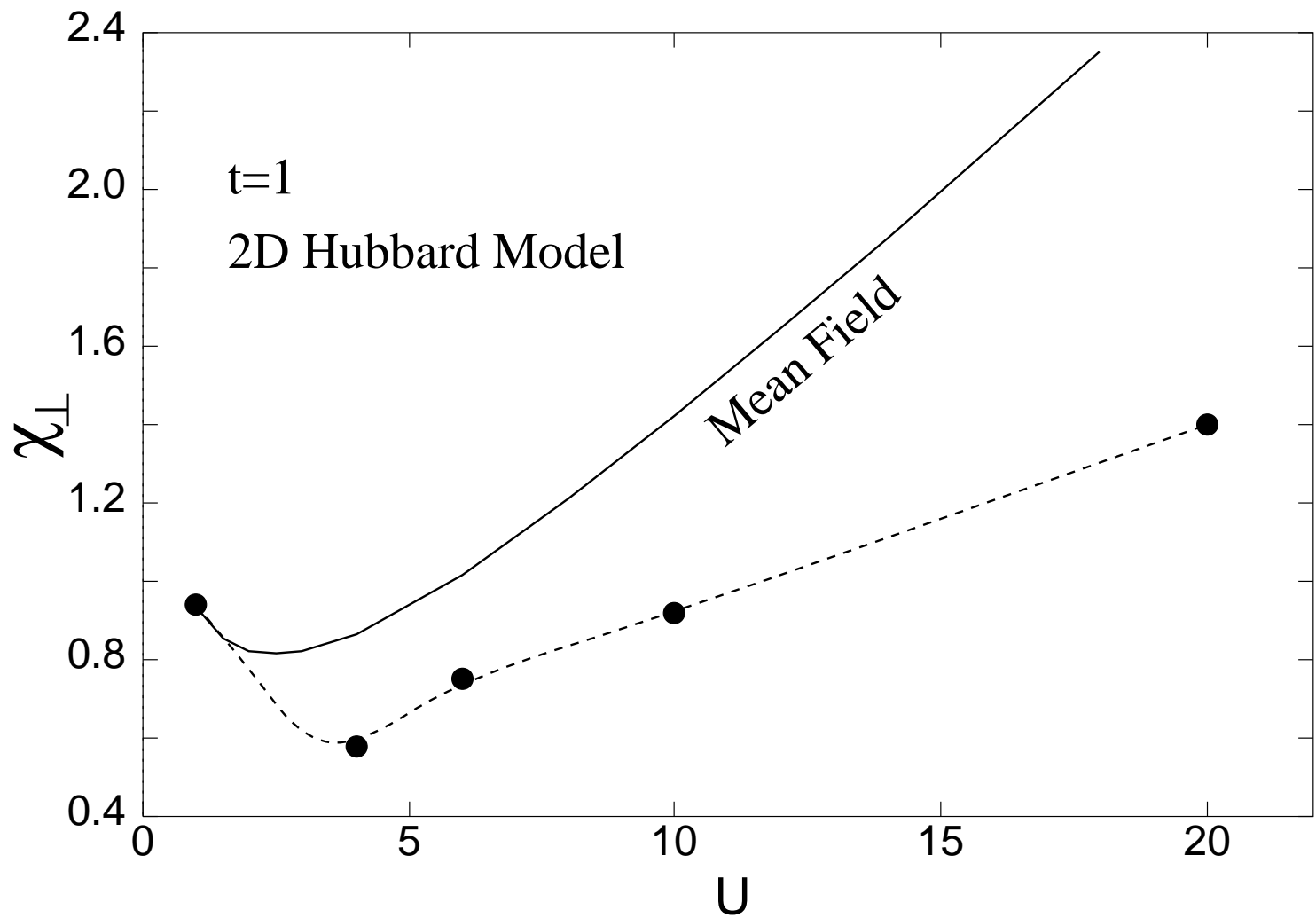
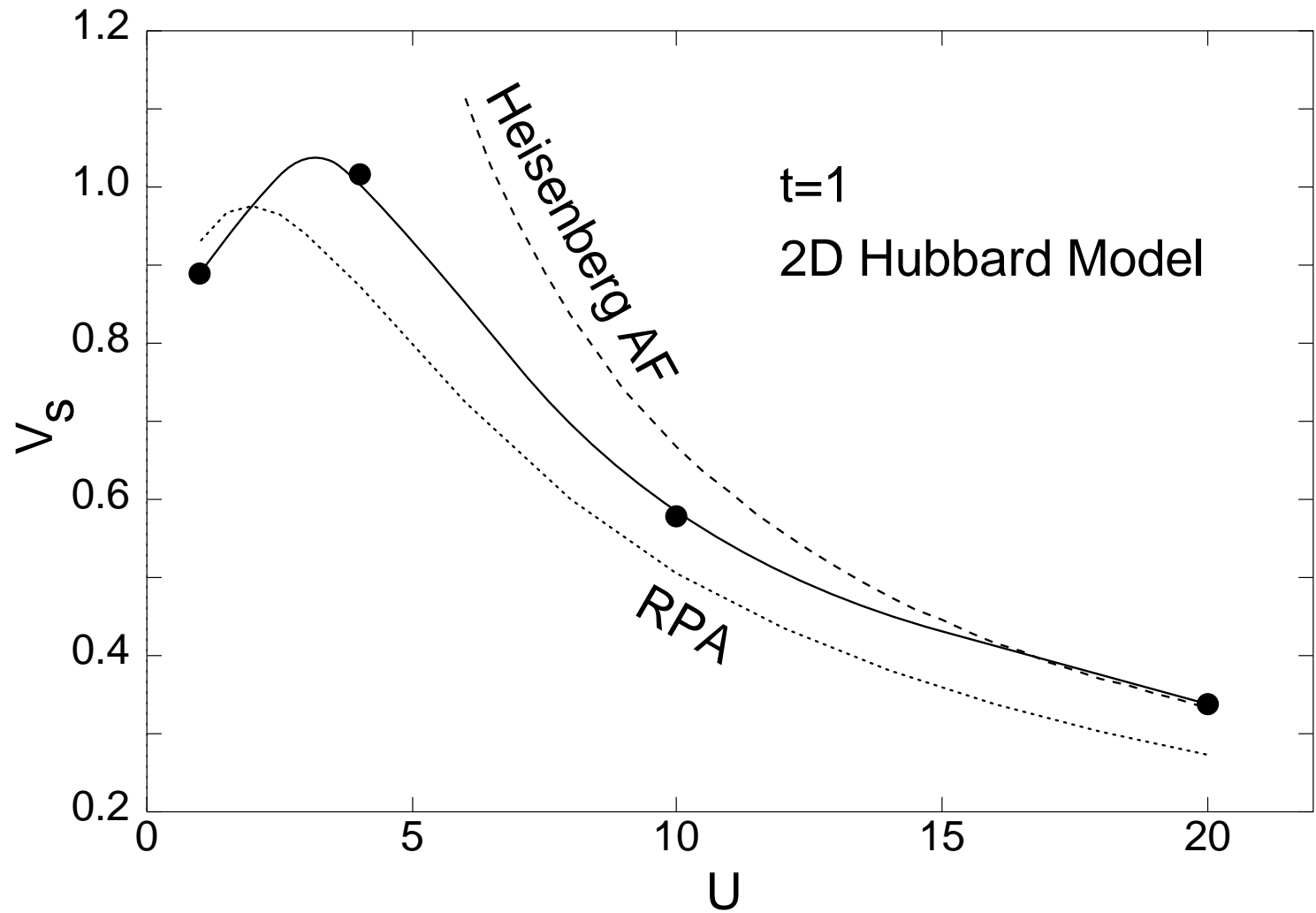
\end{document}